# Dynamics of colloidal crystals studied by pump-probe experiments at FLASH

R. Dronyak[1], J. Gulden[1], O. M. Yefanov[1], A. Singer[1], T. Gorniak[2,3], T. Senkbeil[2,3], J.-M. Meijer[4], A. Al-Shemmary[1], J. Hallmann[5], D. D. Mai[5], T. Reusch[5], D. Dzhigaev[6], R. P. Kurta[1], U. Lorenz[1], A. V. Petukhov[4], S. Düsterer[1], R. Treusch[1], M. N. Strikhanov[6], E. Weckert[1], A. P. Mancuso[7], T. Salditt[5], A. Rosenhahn[2,3], and I. A. Vartanyants[1,6]*

[1] *Deutsches Elektronen-Synchrotron DESY, Notkestraße 85, D-22607 Hamburg, Germany*
[2] *Institute of Functional Interfaces (IFG), Karlsruhe Institute of Technology (KIT), Hermann-von-Helmholtz-Platz 1, 76344 Eggenstein-Leopoldshafen, Germany*
[3] *Angewandte Physikalische Chemie, Ruprecht-Karls-Universität Heidelberg, Im Neuenheimer Feld 253, D-69120 Heidelberg, Germany*
[4] *Van't Hoff Laboratory for Physical and Colloid Chemistry, Debye Institute for Nanomaterials Science, Utrecht University, Padualaan 8, 3584 CH Utrecht, The Netherlands*
[5] *Institut für Röntgenphysik, Georg-August-Universität Göttingen, Friedrich-Hund-Platz 1, D-37077 Göttingen, Germany*
[6] *National Research Nuclear University, "MEPhI", 115409 Moscow, Russia*
[7] *European XFEL GmbH, Albert-Einstein-Ring 19, D-22761 Hamburg, Germany*

## Abstract

We present a time-resolved infrared (IR) pump and extreme-ultraviolet (XUV) probe diffraction experiment to investigate ultrafast structural dynamics in colloidal crystals with picosecond resolution. The experiment was performed at the FLASH facility at DESY with a fundamental wavelength of 8 nm. In our experiment, the temporal changes of Bragg peaks were analyzed and their frequency components were calculated using Fourier analysis. Periodic modulations in the colloidal crystal were localized at a frequency of about 4–5 GHz. Based on the Lamb theory, theoretical calculations of vibrations of the isotropic elastic polystyrene spheres of 400 nm in size reveal a 5.07 GHz eigenfrequency of the ground (breathing) mode.

---

* Corresponding author: Ivan.Vartaniants@desy.de

## Introduction

Artificial hypersonic phononic crystals are a new generation of acousto-optical devices that can be used for ultrafast manipulation and control of electromagnetic waves by hypersonic (GHz) acoustic waves.[1] Colloidal crystals, formed by self-assembly of polystyrene and silica nanospheres have shown phononic band gaps in the GHz frequency range,[2-7] which corresponds to a picosecond time scale. From that respect, probing ultrafast dynamics in colloidal crystals has attracted interest in recent years. There are a few techniques that measure vibrations excited by light waves in nanosized colloidal crystals, such as Brillouin light scattering,[2,5] Raman scattering[8] and optical pump-probe spectroscopy.[4,6,7] However, optical techniques are severely limited in terms of the accessible spatial scales. Moreover, sufficient refractive index matching is often very difficult, if not impossible.

Two- and three-dimensional, static, structural information of colloidal crystals can be measured with high spatial resolution at 3rd generation synchrotron sources using advanced x-ray scattering[9-12] and imaging[13-16] techniques. At the same time, newly developed free-electron lasers (FEL)[17-20] are especially well suited for time-resolved experiments on ultrafast structural dynamics of different materials,[21] including colloids. FELs provide extremely powerful coherent femtosecond x-ray pulses, which are necessary to perform time-resolved experiments with a time resolution that outperforms that of 3rd generation synchrotron sources.

Here, we present results of a time-resolved pump-probe diffraction experiment for investigation of ultrafast structural dynamics of colloidal crystals with picosecond time resolution. In our experiment, the temporal changes in the sample were induced by infrared (IR) laser pulses and could be observed in diffraction patterns produced by the scattered extreme-ultraviolet (XUV) pulses from FLASH after different time delays. Contrary to the pump-probe experiments with visible light, the short wavelength of XUV allows for recording of diffraction patterns which contain Bragg peaks of several orders and the diffuse scattering between them. The scattering data is then free of multiple scattering artifacts due to the intrinsically low scattering contrast of XUV radiation. This gives the possibility of simultaneously extracting the time-dependent variations of the sample in different spatial directions and provides unique information about the dynamics in colloidal samples.

## Experiment

The experiment was conducted at the BL3 beamline at FLASH at a fundamental wavelength of 8 nm. FLASH was operated producing single FEL pulses with an average energy of 170 μJ per pulse at a repetition rate of 10 Hz. The last mirror of the beamline was an ellipsoidal mirror with a focal distance of 2 m, which provides an image of the source. Using laser-induced ablation of polymethylmethacrylate (PMMA) films,[22] the estimated beam size in the sample plane was determined to be 12×40 μm$^2$ (FWHM) in the horizontal and vertical directions, respectively.

The diffraction data were measured in the dedicated HORST vacuum chamber[23], which was connected to the BL3 beamline (Fig. 1). The chamber contains a sample stage which allows for rotation of the sample around the vertical axis, an in-vacuum detector and an optical microscope for alignment of the sample. The FEL beam was incident on the sample at a distance of 70 m from the source. Absorbers were used for attenuation of the intense beam to perform the experiment in the non-destructive regime, where the sample can withstand several shots without detectable damage. A 300 nm thick aluminum foil inserted in the beam reduced the intensity of the beam to $(2.3\pm0.7) \times 10^9$ photons/pulse. The transmitted beam then propagated 174 mm to the detector plane (see Fig. 1), where an in-vacuum charge-coupled device (CCD) (Andor DODX436-BN with 2048×2048 pixels, each with 16-bit digitization and a size of 13.5×13.5 μm$^2$) recorded the data. A cross-shaped beamstop manufactured from stainless steel with the central part covered by $B_4C$ was positioned in front of the CCD to protect the camera from the direct beam (see Fig. 1). A typical single-shot diffraction pattern is shown in Fig. 2, demonstrating the excellent visibility of the Bragg peaks.

In addition to the XUV probe pulses from FLASH, IR laser pulses were used to excite the sample. The pump pulses were generated by a Ti:sapphire laser system[24] operating at 800 nm with a pulse duration of about 100 fs (FWHM) and a pulse energy of 400 μJ focused to a spot size of about 200×200 μm$^2$ in the sample plane. The IR laser pulses were co-propagating along the XUV pulses (see Fig. 1) and synchronized to the pulses from FLASH with less than 500 fs FWHM jitter. The arrival time of the pulses with respect to each other or the time delay Δt, can be varied with an accuracy of better than 500 fs. We measured a series of diffraction patterns with time delays from -100 to +1000 ps, with 100 ps and 50 ps time intervals. For each time delay, ten single pulse diffraction patterns were recorded to increase the statistics.

The pump-probe experiments were performed on a colloidal crystal film. The sample consisted of a Si wafer with an array of 250×250 μm$^2$ sized windows which were sufficiently larger than the FLASH beam. These windows were covered by a 100 nm thick $Si_3N_4$ membrane and coated with the colloidal crystals made of (398±11) nm diameter polystyrene microspheres. The polystyrene spheres were synthesized by emulsifier-free emulsion polymerization of styrene using potassium persulfate as an initiator.[25] Colloidal crystals were grown by the vertical deposition technique.[26,27] For this, the wafer was immersed in a suspension of spheres in ethanol

(p.a.) with a volume fraction of 0.1 vol.% in a temperature-controlled room at 20º C. The sample thickness was about 11 layers (see Figs. 1(a) and 1(b)).

The measurements were performed in two different orientations of the sample. The initial orientation with the azimuthal angle $\Delta\varphi = 0º$ (see Fig. 1) corresponds to a direction of the incident IR and XUV pulses along the [111] direction of the fcc colloidal crystal, which is perpendicular to the sample surface. In this geometry we were probing dynamics parallel to the surface of the crystal. The second orientation was chosen at the incident angle $\Delta\varphi = 35º$ (not shown), where another set of Bragg peaks can be accessed. This direction corresponds to the [110] crystallographic fcc direction and the experiment is sensitive to the dynamics induced in the sample both along and perpendicular to the film surface. Our analysis has shown that in the conditions of our experiment we obtained similar results in both geometries.

Figure 2 shows a typical single-shot diffraction pattern measured at $\Delta\varphi = 0º$ after background subtraction. The background signal from the pump laser was removed by averaging over ten images recorded without the probe pulse and subtracting the result from all diffraction patterns. Several orders of the Bragg peaks can be well observed on a single-shot diffraction pattern as shown in Fig. 2. The three strongest families of reflections, which were used in further analysis, are marked. The family of peaks marked as ($\bar{2}\bar{2}0$) in Fig. 2 are the $\bar{2}\bar{2}0$ fcc reflections. The other two sets of reflections 1/3(422) and 2/3(422) are forbidden in a perfect fcc crystal. They reflect the periodicity within a single close-packed layer of colloidal spheres, which are visible due to the small sample thickness and random stacks of hexagonal close-packed planes.[27] The shape of each Bragg peak is broadened in both vertical and horizontal directions. Furthermore, the speckles are well resolved, due to coherent illumination of the sample. Notice also the diffuse scattering around and between the Bragg peaks.

## Results and discussion

When the energy of the IR pump pulse is transferred to the crystal, changes in the colloidal spheres induce crystal lattice dynamics that affect the position, intensity and shape of Bragg peaks.[28] To reveal the characteristic time scales on which structural changes occur in the colloidal crystal during the measurement, we have analyzed each single-shot diffraction pattern in the following way: We selected three sets of reflections as shown in Fig. 2 for analysis and calculated their center of mass starting from an initial guess. Then, the center of each diffraction pattern and the relative peak to center distances that correspond to the momentum transfer vector $\mathbf{Q}$ (see Fig. 2) for each time delay $\Delta t$ were determined. In this way the lattice dynamics in the relevant directions in the crystal was studied. To characterize the changes of the peak shapes, each of them was fitted by a two-dimensional Gaussian function giving its size (FWHM) both in the horizontal ($W_x$) and vertical ($W_y$), directions (see Fig. 2). The time dependent broadening of the Bragg peaks corresponds to an inhomogeneous disorder in the crystal lattice induced by the pump laser. To characterize the changes in the relative peak to background intensities, which would also correspond to the modulation of the crystal lattice spacing, we calculated the ratio of the integrated intensity of the peak to the diffuse scattering in a region between the peaks. For each time delay $\Delta t$, all parameters described above were averaged over ten diffraction patterns.

In Fig. 3 the time-dependence of the momentum transfer vector **Q**, and horizontal $W_x$ and vertical $W_y$ size of the selected 2/3(422) and $\overline{2}\overline{2}0$ peaks (see Fig. 2) are presented, as an example. To estimate characteristic frequencies probed in the experiment the Fourier transforms of all time-dependent parameters were calculated. The corresponding power spectrum[29] is shown in the insets of Fig. 3. Our results indicate an increased contribution of the Fourier components in the frequency region of 4–5 GHz.

To determine the mechanism underlying the dynamics observed in the pump-probe experiment we compare our results with the Lamb theory[30]. This theory is based on the analysis of vibrations of an isotropic elastic sphere. In general, the vibrations of an elastic sphere can be classified by its spheroidal and torsional modes. Spheroidal modes lead to the change of volume of the sphere due to the radial displacement, while the torsional modes involve only shear motions for which the volume of the sphere stays constant. In optical pump-probe experiments the change in the volume of spheres can be probed. Here, we limit our analysis to the spheroidal modes which can be classified in terms of the order $l$ ($l = 0, 1, 2, \ldots$) of the spherical Bessel functions and the number of nodes $n$ ($n = 1, 2, 3, \ldots$) in the radial directions. In our calculations we used a sphere diameter $d = 400$ nm and the values of the longitudinal $c_L = 2350$ m/s and transverse $c_T = 1210$ m/s sound velocities in polystyrene.[31] For the ground mode ($n = 1$, $l = 0$) the displacement is purely radial and the calculated eigenfrequency of this breathing mode is 5.07 GHz which corresponds to the period of vibrations of 197 ps. This analysis shows that the theoretical frequency of the breathing mode is within the range of the experimentally observed values.

## Summary

In summary, we measured the diffraction patterns from colloidal crystals, which were pumped by short IR laser pulses and probed by FEL radiation while changing the time delay between the lasers in the picosecond range. The studies were carried out in a non-destructive regime at different rotation angles of the sample. The changes in the colloidal crystal induced by the IR laser were investigated via analysis of Bragg peaks extracted from diffraction patterns. The dynamics at different time scales were studied through Fourier analysis of parameters associated with the momentum transfer and horizontal and vertical peak size. An enhancement of the spectrum in a frequency range of about 4–5 GHz was observed. Theoretical calculations of vibrations based on the Lamb theory of a 400 nm size polystyrene sphere used in the experiment, reveal an eigenfrequency of the ground (breathing) mode in the same frequency range. This analysis suggests that the same modes were excited in our pump-probe experiment. However, to give a definitive answer to the observed dynamics induced in colloidal crystals by the propagation of IR pulses it will be desirable in future experiments to increase the shot-to-shot stability of the incoming FEL pulses, which could significantly reduce associated errors in the measurements. With the future seeded FEL sources[32] it should become possible to achieve much more stable conditions and consequently to reach higher statistics on the extracted parameters of structural dynamics.

The demonstrated pump-probe experiments combined with femtosecond coherent x-ray diffraction imaging technique[33-36] have the potential to visualize ultrafast dynamics in colloidal crystals with nanometer spatial resolution at femtosecond time scales.

Part of this work was supported by BMBF Proposal 05K10CHG "Coherent Diffraction Imaging and Scattering of Ultrashort Coherent Pulses with Matter" in the framework of the German-Russian collaboration "Development and Use of Accelerator-Based Photon Sources". We acknowledge funding by the BMBF project 05K10VH4 within the FSP 301 FLASH and the Virtual Institute VH-VI-403 of the Helmholtz association.

# Figure captions

**Figure 1.** Schematic view of the pump-probe experiment showing the infrared pump (IR pulse) and the extreme-ultraviolet probe (FEL pulse) separated by a time delay Δt, the sample, and the detector which is protected by the beamstop. Insets (a) and (b) show SEM images of the colloidal crystal film used in the experiment. The eleven layers of polystyrene spheres composing the colloidal crystal are visible in the inset (b).

**Figure 2.** Selected single-shot diffraction pattern on a logarithmic scale measured in transmission geometry at azimuthal angle orientation Δφ = 0°. The family of fcc peaks ($\overline{22}0$) and forbidden fcc peaks (1/3(422) and 2/3(422)) are indicated. The momentum transfer vector **Q** and the horizontal $W_x$ and vertical $W_y$ size of the peaks are shown.

**Figure 3.** Time dependence of the momentum transfer **Q** (a, d), horizontal $W_x$ (b, e) and vertical $W_y$ (c, f) size (FWHM) for the selected Bragg peaks 2/3(422) (a-c) and $\overline{22}0$ (d-e) shown in Fig. 2. Error bars are determined as a standard deviation for ten measurements. The insets show the power spectrum of the corresponding data.

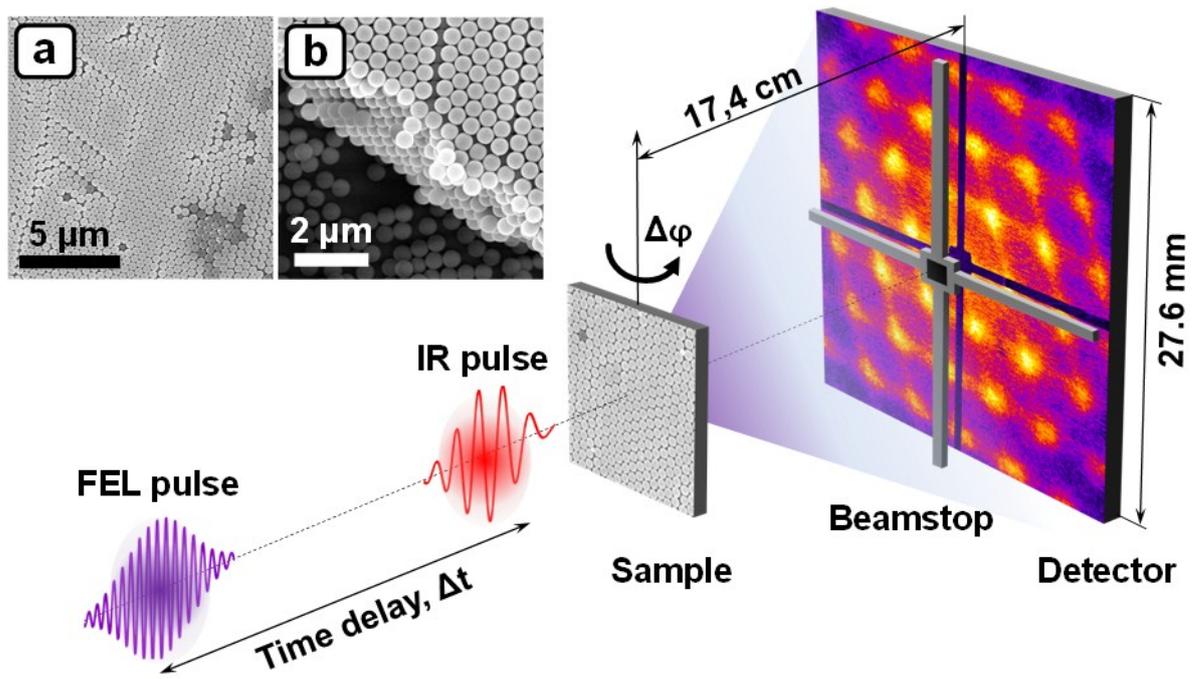

**Figure 1**

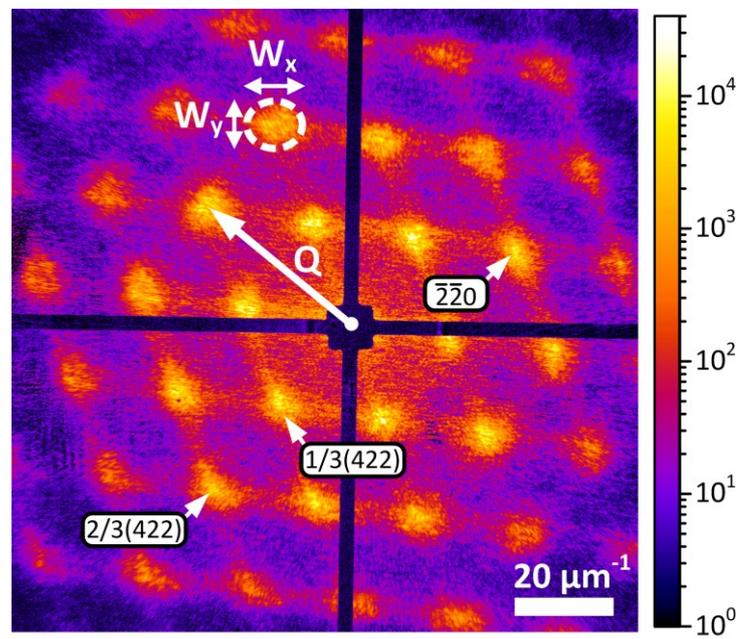

**Figure 2**

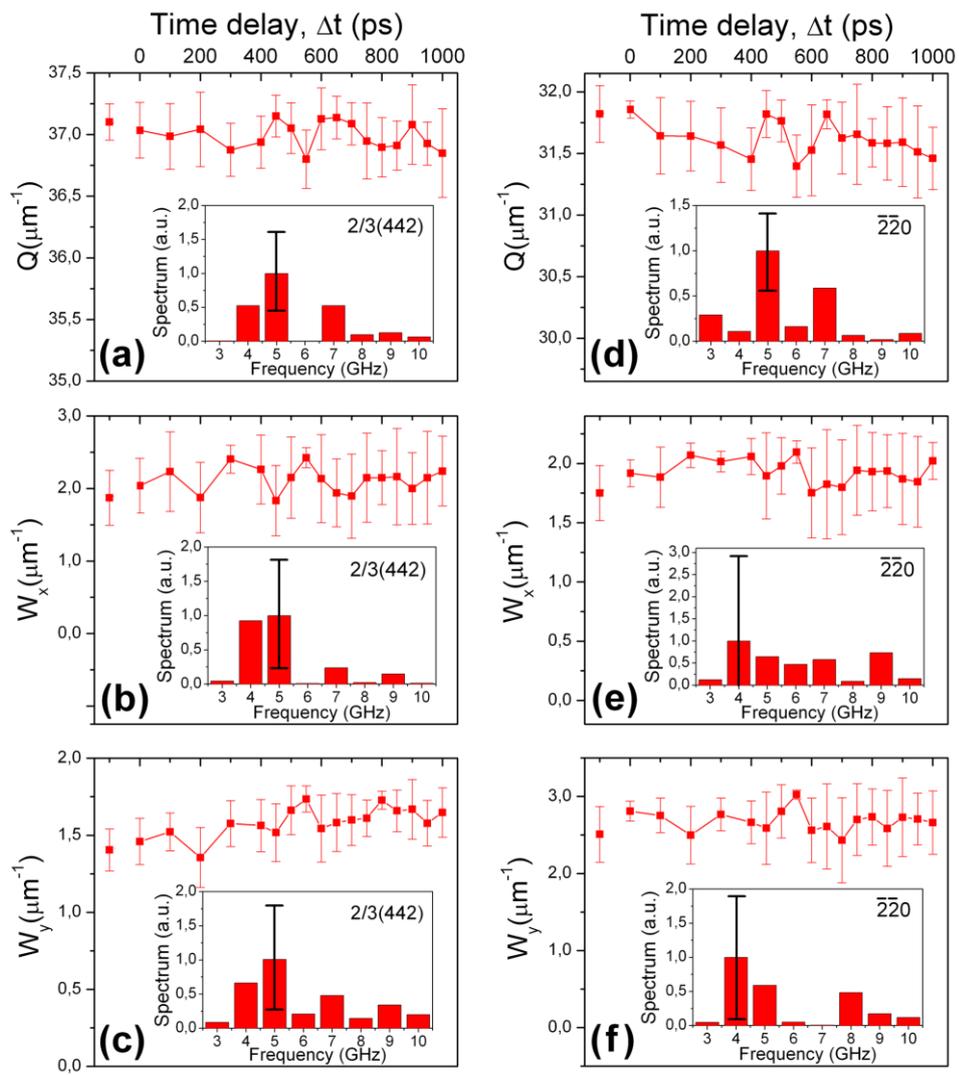

**Figure 3**